\documentclass[reprint,amsmath,amssymb,aps,]{revtex4-2}

\usepackage{graphicx,float,color,multirow,dcolumn,ulem,comment,hyperref}
\hypersetup{
colorlinks=true,
linkcolor=blue,
urlcolor=blue,
citecolor=blue}

\makeatletter
\newcommand{\Rmnum}[1]{\expandafter\@slowromancap\romannumeral #1@} % 大写罗马数字

\makeatother

\begin{document}

\title{Critical phenomenon inside asymptotically flat black holes with spontaneous scalarization} 

%- authors and addresses -
\author{Li Li$^{1,2,3}$}\email{liliphy@itp.ac.cn}
\author{Ze Sun$^{1,2,3}$}\email{sunze23@mails.ucas.ac.cn}
\author{Fu-Guo Yang$^{3,4}$}\email{yangfuguo@ucas.ac.cn}

%\affiliation{}
\affiliation{$^1$School of Fundamental Physics and Mathematical Sciences, Hangzhou Institute for Advanced Study, UCAS, Hangzhou 310024, China.}

\affiliation{$^2$Institute of Theoretical Physics, Chinese Academy of Sciences, P.O. Box 2735, Beijing 100190, China.}

\affiliation{$^3$School of Physical Sciences, University of Chinese Academy of Sciences, Beijing, 100049, China}

\affiliation{$^4$ International Centre for Theoretical Physics Asia-Pacific, University of Chinese Academy of Sciences, 100190 Beijing, China}

\begin{abstract}

We study the nonlinear interior dynamics of spontaneously scalarized black holes in Einstein–Maxwell–scalar theory with zero cosmological constant, revealing novel critical phenomena. We demonstrate that, for a wide range of scalar-electromagnetic couplings, scalarized black holes possess no smooth inner Cauchy horizon and instead evolve into a spacelike Kasner singularity. The scalar hair triggers a rapid collapse of the Einstein-Rosen bridge at the would-be Cauchy horizon. Near the critical point where scalarized black holes bifurcate from the Reissner–Nordström solution, we establish a robust scaling relation between the Kasner exponent and the charge-to-mass ratio of the hairy black hole, opening a new window into the remarkable simplicity underlying black hole interiors.

\end{abstract}

\maketitle

\textbf{Introduction.}--Critical phenomena in nature reveal how vastly different systems—from boiling water to collapsing stars—can exhibit universal behavior near a tipping point, where characteristic scaling laws emerge that are independent of the system's microscopic details. Critical phenomena have been uncovered in black hole physics, including black hole thermodynamics and critical collapse. In the former, critical points—such as in charged Anti-de Sitter black holes~\cite{Mann:2025xrb}—demonstrate phase transitions with precise scaling laws and universal critical exponents, directly linking gravitational physics to statistical mechanics and holographic duality. In gravitational collapse, Choptuik's discovery of mass-scaling and discrete self-similarity at the black hole formation threshold exposes a deep, scale-invariant structure in Einstein's equations, where black holes form according to power-law relations independent of initial conditions, see~\cite{Gundlach:2002sx} for a review. These phenomena underscore that gravity, despite its nonlinear complexity, obeys universal laws near critical points, offering crucial insights into quantum gravity, cosmic censorship, and the fundamental nature of spacetime singularities. However, these investigations have primarily focused on the exterior of black holes. In recent years, significant attention has been directed toward the internal dynamics of black hole with various hairs, revealing far more complex and rich behaviors than those observed externally, such as the non-existence of inner horizons~\cite{Cai:2020wrp,An:2021plu}, Einstein-Rosen (ER) bridge collapse~\cite{Hartnoll:2020rwq}, Josephson oscillations for scalar/vector fields~\cite{Hartnoll:2020fhc,Cai:2021obq}, and alternation between Kasner epochs~\cite{Cai:2023igv,Cai:2024ltu,Hartnoll:2022rdv}. It is thus an intriguing and important question whether one can uncover critical phenomena within black holes.

A particularly interesting case involves spherical black holes in Einstein-Maxwell-scalar (EMS) models  with vanishing cosmological constant~\cite{Herdeiro:2018wub,Fernandes:2019rez}, where a massless scalar field couples to the electromagnetic invariant. A physical motivation to study these models is that they arise naturally in Kaluza-Klein and String theories~\cite{Gibbons:1987ps,Lu:2019jus}. Tachyonic instabilities can trigger spontaneous scalarization of the Reissner–Nordström (RN) solution, leading to the formation of thermodynamically stable scalarized black holes and altering the spacetime geometry (see~\cite{Doneva:2022ewd} for a recent review on spontaneous scalarization). Consequently, scalar hair can leave observable imprints—such as modifications in black hole shadows~\cite{Gan:2021xdl,Al-Badawi:2024dzc,Wu:2025hcu}—offering opportunities to test the no-hair theorem and explore alternative theories of gravity. While the exterior properties of these scalarized black holes have been extensively investigated, their interior structure and dynamics remain largely unexplored.

In this work, we uncover the nonlinear nature of the dynamics inside asymptotically flat black holes with spontaneous scalarization in EMS theories. For a broad class of models, we demonstrate that scalarized black holes have no smooth inner horizon and terminate in a Kasner singularity. Moreover, the collapse of the ER bridge—initiated by a neutral scalar field at the would-be inner horizon—is an intrinsic feature. Near the critical point where scalarized black holes bifurcate from the RN solution, we identify a robust scaling relation between the Kasner exponent and the charge-to-mass ratio of the hairy black hole, revealing deep universal principles that govern general relativity in its most extreme regimes. Throughout this work, we adopt units in which $G=c=1$.

\textbf{Model and scalarized black holes.}--The action in 4-dimensional EMS theory reads
\begin{equation}\label{Action}
\begin{split}
S&=\frac{1}{16\pi}\int d^{4}x \sqrt{-g}[\mathcal R+\mathcal L_{M}]\,,\\
\mathcal L_{M}&=-2\partial_\mu \psi \partial^\mu \psi -Z(\psi) F_{\mu\nu}F^{\mu\nu}\,,
\end{split}
\end{equation}
where $\mathcal R$ is the Ricci scalar curvature, $\psi$ is a real scalar, and $F_{\mu\nu}=\nabla_{\mu} A_{\nu}-\nabla_{\nu} A_{\mu}$ is the field strength of the gauge potential $A_\mu$. To recover the Einstein-Maxwell case in the absence of $\psi$, the scalar-gauge coupling function must satisfy $\frac{dZ}{d\psi}|_{\psi=0} = 0$. We choose $Z(0) = 1$ without loss of generality. %\textcolor{red}{Therefore, the general form of the coupling function expansion is $1+\alpha^2 \psi^2+\cdots$, where $\alpha^2>0$ is a constant representing the coupling strength.}

We begin with the static, spherically symmetric black holes with a purely electric field in asymptotically flat spacetime. The ansatz is therefore given by
\begin{equation}\label{Ansatz}
\begin{aligned}
ds^2 = \frac{1}{z^2} \left[- f(z) \mathrm{e}^{-2\chi(z)} dt^2 + \frac{dz^2}{f(z)} +d\Omega^2_2 \right],\\
\psi=\psi(z),\quad A_{\mu}{dx^\mu}=A_t(z)dt\,,
\end{aligned}
\end{equation}
with $d\Omega^2_2\equiv d\theta^2+\sin^2\theta d\varphi^2$ the metric of a unit 2-sphere. Moreover, the functions $f,\chi, \psi, A_t$ are all radially dependent only. We denote the position of the event horizon as $z_H$ at which the blackening function $f(z_H)=0$. The area of the event horizon is $A_H=4\pi/z_H^2$. 
The asymptotic region lies at $z=0$, with the interval $0<z<z_H$ corresponding to the black hole exterior. The interior region, extending from the event horizon to the singularity, is described by $z_H<z<\infty$. Without loss of generality, we work in a gauge where $A_t$ vanishes on the event horizon. The smoothness of the geometry at the horizon admits the Taylor expansion for all fields at $z_H$. At the spatial infinity $z=0$, one imposes $\chi(0) = \psi(0) = 0$ with
\begin{equation}\label{z0BoundaryConditions}
\begin{aligned}
\frac{f(z)}{z^2} =1-2Mz+\cdots,\; A_t(z)=\mu - Q z + \cdots\,,
\end{aligned}
\end{equation}
where $M$, $Q$ and $\mu$ denote the ADM mass, electric charge and electrostatic potential, respectively. 

\begin{figure}[H]
\includegraphics[width=0.45\textwidth]{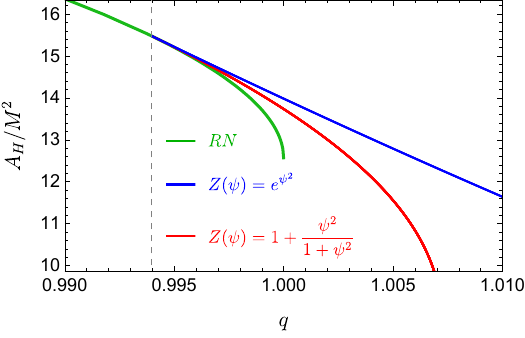}
\caption{An illustration of spontaneously scalarized black holes under two representative coupling functions, showing the reduced event horizon area $A_H/ M^2$ as a function of the charge-to-mass ratio $q$. The scalarized black holes bifurcate smoothly from the RN solution (green curve) at the critical charge-to-mass ratio $q_c$ (vertical dashed line). }\label{FigPhaseAHQ} 
\end{figure}

For given coupling function $Z(\psi)$, there is a one-parameter family of hairy solutions, which can be labeled by the charge-to-mass ratio $q=Q/M$. For more details, please refer to the Supplementary Material (SM)~\footnote{The Supplementary Material [url], which includes Refs.~\cite{Astefanesei:2019pfq,Herdeiro:2018wub,Sword:2021pfm,Cai:2023igv,Belinskii:1973sud,Kasner:1921,Dias:2021afz,EventHorizonTelescope:2019ths,EventHorizonTelescope:2019kwo,Gan:2021xdl,Al-Badawi:2024dzc,Wu:2025hcu,Gralla:2019a,Chael:2021inner}, presents the technical details for solving the equations of motion, analyzing interior dynamics, and computing black hole images, thereby supporting the results made in the main text. }. When the scalar field vanishes, the only charged black hole from~\eqref{Ansatz} is given by the RN solution which reads
\begin{equation}\label{eqRN}
\begin{aligned}
f(z)&=z^2(1-2 M z +Q^2z^2)\,,\\
A_t(z)&=\mu - Q z=Q(z_H-z) \,,
\end{aligned}
\end{equation}
together with $\chi(z)=\psi(z)=0$. Besides the event horizon $z_H=(M-\sqrt{M^2-Q^2})/{ Q^2}$, there is a Cauchy horizon at $z=z_I=(M+\sqrt{M^2-Q^2})/{ Q^2}$ inside the RN black hole. 

The two representative hairy black holes are depicted in Fig.~\ref{FigPhaseAHQ}. While both share the same small-$\psi$ expansion $1+\psi^2+\cdots$, they exhibit significantly different behavior in the large-$\psi$ limit. It is evident from Fig.~\ref{FigPhaseAHQ}, at the critical charge-mass ratio $q_c$, the scalarized black holes bifurcate from RN black holes (green curve), and reduce to the latter for $\psi=0$. This bifurcation is associated to a tachyonic instability of the RN black hole~\eqref{eqRN}. Furthermore, for a given $q$, the hairy black hole possesses a larger event horizon area—and thus greater entropy—making it thermodynamically favored from the viewpoint of black hole thermodynamics.

\textbf{Interior structure.}--While the RN black hole~\eqref{eqRN} has an inner Cauchy horizon, we prove that the scalarized black holes cannot possess smooth inner horizons when 
\begin{equation}\label{NoHorizonCondition}
    \psi \frac{dZ}{d\psi}>0\,.
\end{equation}
Suppose that there is an inner horizon $z_I$ inside the event horizon $z_H$ for a scalarised black hole. The function $f(z)$ is vanishing at both horizons and is negative between the two horizons. The integral of the equation of motion for $\psi$ implies that
\begin{equation}\label{IntegralPsi}
\begin{aligned}
0&=\int_{z_\text{H}}^{z_I} \left(z^{-2}\mathrm{e}^{-\chi}f\psi'\psi\right)'dz\\
&=\int_{z_\text{H}}^{z_I}\left(z^{-2}\mathrm{e}^{-\chi}f\psi'^2-\frac{1}{2}\mathrm{e}^{\chi}A_t'^2\psi\frac{dZ}{d\psi}\right)dz\,.
\end{aligned}
\end{equation}
Under~\eqref{NoHorizonCondition}, the integrand in the second line is non-positive over the range of integration. Therefore, the scalar field necessarily removes the inner horizon, and thus the scalarised black hole ends at a spacelike singularity as $z\rightarrow\infty$. This finding demonstrates that the Cauchy horizon instability inherent to strong cosmic censorship extends to scalarized branches, thus restoring predictability for a more extensive family of charged black hole interiors.
We highlight that in order to trigger a spontaneous scalarisation, the coupling should satisfy the condition~\eqref{NoHorizonCondition} for some range of $z$ outside the black hole~\cite{Astefanesei:2019pfq}. In practice, the condition~\eqref{NoHorizonCondition} applies to a wide class of coupling functions commonly used in EMS models. Note that the conserved charge method~\cite{Cai:2020wrp} for removing the inner horizon does not apply to our system.

Consequently, when $q$ slightly exceeds the critical value $q_c$, the scalar field triggers an instability of the inner Cauchy horizon in the RN background. A phenomenon widely observed in previous studies~\cite{Hartnoll:2020rwq,Hartnoll:2020fhc,Dias:2021afz,Cai:2021obq,Xu:2025edz,Liu:2022rsy,Grandi:2021ajl} is the rapid decay of the $g_{tt}$ component as it approaches its would-be zero value at the Cauchy horizon, \emph{i.e.} the collapse of the ER bridge. However, it was recently argued that this collapse could be fully suppressed by strengthening the EMS coupling~\cite{Sword:2021pfm}. We show that this suppression arises because the state is not sufficiently close to the critical value. By tuning the parameter toward its critical value, a clear collapse of the ER bridge should be restored (see Appendix A).

Closely following the ER bridge collapse, a robust outcome during the late-time interior evolution is the emergence of a Kasner geometry, for which the solution is approximately given by
\begin{equation}\label{ApproximateSolution}
\psi=\beta \ln z+C_{\psi}\,,\ \chi =\beta^2\ln z +C_{\chi}\,,\ f=C_f z^{3+\beta^2}\,, 
\end{equation}
with $\beta$, $C_{\psi}$, $C_{\chi}$ and { $C_{f}$} constants. The Kasner structure is manifest by considering the proper time $\tau\sim z^{-\frac{3+\beta^2}{2}}$:
\begin{equation}\label{eqKasner}
\begin{aligned}
\mathrm{d}s^2=-\mathrm{d}\tau^2+\tau^{2p_t}\mathrm{d}t^2+\tau^{2p_s}\mathrm{d}\Omega^2_{2},\;\;
\psi= -\frac{p_{\psi}}{\sqrt{2}}\ln \tau\,,
\end{aligned}
\end{equation}
with the Kasner exponents
\begin{equation}\label{pts}
p_t=\frac{\beta^2-1}{\beta^2+3},\ p_s=\frac{2}{\beta^2+3},\ p_\psi=\frac{2\sqrt{2}\beta}{\beta^2+3}\,,
\end{equation}
satisfying two constraints:
\begin{equation}
  p_t+2p_s=1,\quad p_t^2+2 p_s^2+p_{\psi}^2=1\,.
\end{equation}
As $\tau\rightarrow 0$ ($z\rightarrow\infty$), the spacetime curvature diverges, signaling a spacelike singularity. This is called the Kasner singularity with the Kasner exponents fully characterized by the parameter $\beta$~\cite{Belinskii:1973sud,Kasner:1921}. For the EMS theory~\eqref{Action}, we find a stable  Kasner geometry all the way to the singularity, in contrast to the rich alternation of Kasner epochs for a charged scalar $\Psi$ due to the  new interaction $\sim A_\mu A^\mu |\Psi|^2$ and non-trivial scalar potential~\cite{Cai:2020wrp,Cai:2023igv} (see SM for more details).

\begin{figure}[h!]
\includegraphics[width=0.45\textwidth]{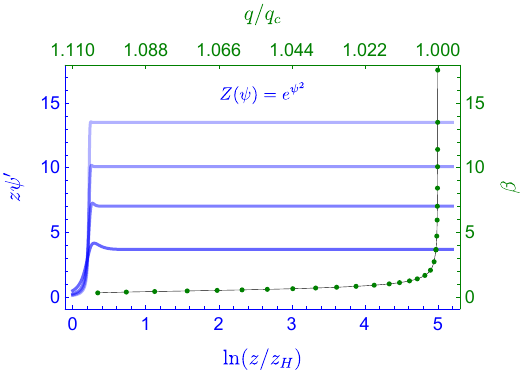}\\
\includegraphics[width=0.45\textwidth]{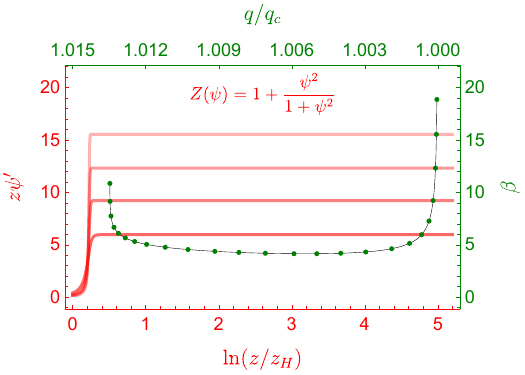}
\caption{The interior dynamics of the scalar field for $Z=e^{\psi^2}$ (top panel) and $Z=1+\psi^2/(1+\psi^2)$ (bottom panel). The plateaus correspond to Kasner geometries, for which $(q/q_c-1)$ for the successive plateaus (from top to bottom) are $\{0.33, 0.60, 1.28, 5.38\} \times 10^{-4}$ in the top panel, and $\{0.56, 0.94, 1.89, 6.64\} \times 10^{-4}$ in the bottom panel. Kasner parameter $\beta$ versus $q$ is denoted by the dotted green curves.}\label{FigPhaseBeta}
\end{figure}

\textbf{Critical phenomenon.}--We have shown that $\psi$ grows logarithmically in a Kasner epoch, as evidenced by the plateaus in Fig.~\ref{FigPhaseBeta}. As approaching $q_c$, the plateau height increases, corresponding to larger values of exponent $\beta$. The dependence of $\beta$ on $q$, represented by the green dotted curves in Fig.~\ref{FigPhaseBeta}, varies with the choice of coupling function $Z$. For $Z=e^{\psi^2}$, $\beta$ decreases monotonically with increasing $q$ and can fall below unity. In contrast, for $Z=1+\psi^2/(1+\psi^2)$ in the bottom panel of Fig.~\ref{FigPhaseBeta}, as $q$ is increased, $\beta$ initially decreases monotonically until a minimum is attained, after which it increases. 

\begin{figure}[h!]
\includegraphics[width=0.44\textwidth]{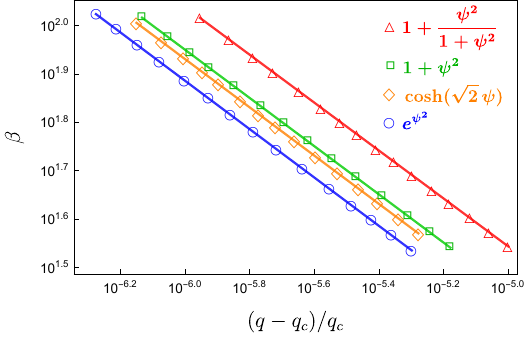}\\
\includegraphics[width=0.44\textwidth]{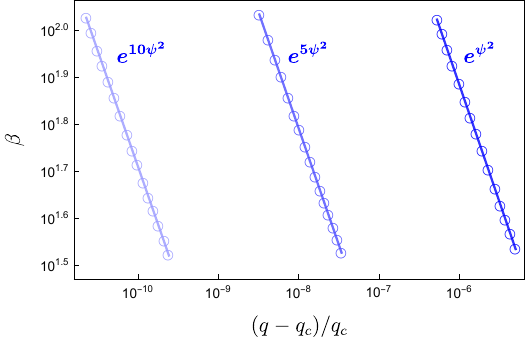}
\caption{The critical behavior between the Kasner parameter $\beta$ and the charge-to-mass ratio $q$. \textbf{Top panel}: Results for different coupling functions.  \textbf{Bottom panel}: Dependence on the coupling strength $\alpha^2$ for $Z=e^{\alpha^2\psi^2}$. Symbols of varying shapes represent numerical data, while the solid line corresponds to the scaling relation~\eqref{BetaToQ}. }\label{FigCriticalBetaQ}
\end{figure}

These rich phenomena are fundamentally rooted in the nonlinear dynamics inside black holes. While the overall trend of $\beta$ versus $q$ depends on the specific form of $Z(\psi)$, a common feature emerges across models: $\beta$ exhibits a sharp increase near $q_c$ and appears to diverge (dotted green curves in Fig.~\ref{FigPhaseBeta}). This consistent divergence pattern suggests the possible presence of universal scaling behavior in the vicinity of the critical point, which we now proceed to investigate. Based on a careful analysis of the extensive numerical data near the critical point, we discover the following scaling law:
\begin{equation}\label{BetaToQ}
\beta= c_{0}\left(\frac{q}{q_c}-1\right)^{-\gamma}, \quad \gamma=0.5\,.
\end{equation}
Here $c_0$ is a constant that depends on the specific details of the model. In contrast, the critical exponent $\gamma=0.5$ is model-independent. As demonstrated in the upper panel of Fig.~\ref{FigCriticalBetaQ}, the scaling law~\eqref{BetaToQ} persists across several different couplings. While the ER bridge collapse is apparently suppressed by strengthening the EMS coupling at fixed $q/q_c$ (top panel of Fig.~\ref{FigCauchyHorizon}), the scaling behavior remains robust. As shown in the bottom panel of Fig.~\ref{FigCriticalBetaQ} for $Z=e^{\alpha^2\psi^2}$, this universality holds across different values of the coupling strength $\alpha^2$.

Given the complexity of black hole interior dynamics, the discovery of a universal scaling behavior appears surprising. Analytically understanding this behavior poses a significant challenge, as nonlinear effects are predominant. Nevertheless, we notice that when the spacetime enters the  Kasner epoch near $q_c$, the scalar field is still weak. For all cases we have numerically checked, the coupling functions share the same small-$\psi$ expansion
\begin{equation}\label{eqZ}
Z(\psi) \sim 1 +  \alpha^2\, \psi^{2}+\cdots\,,
\end{equation}
with $\alpha$ constant. It is known as scalarized-connected-type models for which the scalarized black holes bifurcate from RN black holes and reduce to the latter for $\psi=0$.
Therefore, we conjecture that all scalarized-connected-type models could yield the same critical exponent.  

In terms of the Kasner exponents~\eqref{pts}, one obtains the following scaling:
\begin{equation}\label{exponent}
 p_t\sim(q-q_c)^{\gamma_t}, \;\;  p_s\sim(q-q_c)^{\gamma_s}, \;\;
 p_\psi\sim(q-q_c)^{\gamma_\psi}\,,
\end{equation}
with the critical exponents $\gamma_t=0$ and $\gamma_s=2\gamma_\psi=1$. This likely represents a universal class of Kasner criticality. If the coupling $Z(\psi)$ admits a small-$\psi$ expansion different from~\eqref{eqZ}, the resulting scaling may belong to a distinct universality class. Nevertheless, whenever $\beta$ diverges near $q_c$, the critical exponents in~\eqref{exponent} consistently satisfy the relations $\gamma_t=0$ and $\gamma_s=2\gamma_\psi$. This is in contrast to the uncharged black hole where $\beta$ is continuous and vanishing at the critical point~\cite{Liu:2021hap}. Furthermore, we clarify why the critical scaling behavior~\eqref{BetaToQ} or~\eqref{exponent} is not observed in the charged scalar scenario. The key reason lies in the presence of Josephson oscillations in the complex scalar field $\Psi$, which causes the Kasner parameter $\beta$ to scale proportionally to an oscillatory function of the system parameters. It was numerically found that near $q_c$, $\beta\sim \sin[\frac{a}{q-q_c}+b]$ with $a$ and $b$ constants~\cite{Dias:2021afz}. Moreover, such strong oscillations lead to intervals in which the Kasner epoch becomes unstable, triggering subsequent Kasner transitions in the interior evolution, see more examples in Anti-de Sitter  spacetime~\cite{Cai:2020wrp,Cai:2023igv,Hartnoll:2020fhc}

Spontaneous scalarization can be understood as a continuous phase transition described by the phenomenological Landau model. A recent study on scalarized neutron stars in scalar-tensor theory~\cite{Muniz:2025egq} showed that the critical exponents align with Landau theory's predictions, and a Landau-type ansatz accurately fits the free energy near the critical point. Interestingly, we demonstrate that a similar picture applies to black holes, with critical exponents fully consistent with Landau's theory (see Appendix B for more details). The phase transition picture is based on the geometry outside the event horizon, which near the critical point $q_c$ remains nearly identical to the RN solution and highly smooth. This stands in stark contrast to the interior geometry, which undergoes drastic change near $q_c$. Our work thus uncovers a distinct critical scaling inside the horizon, a novel interior manifestation of criticality not addressed in previous studies. The origin of this interior criticality lies in the highly nonlinear dynamics inside black holes, presenting a key open question for future investigation.

\begin{figure}[h!]
\includegraphics[width=0.15\textwidth]{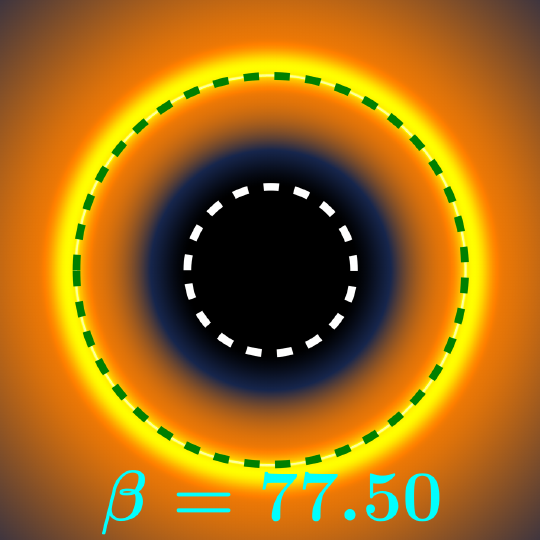}
\includegraphics[width=0.15\textwidth]{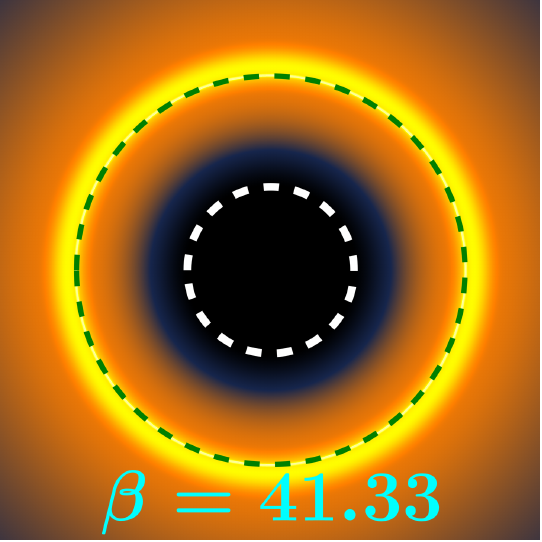}
\includegraphics[width=0.15\textwidth]{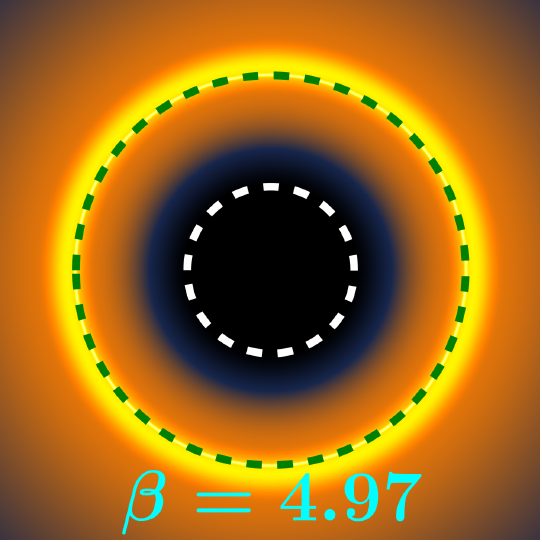}\\
\includegraphics[width=0.15\textwidth]{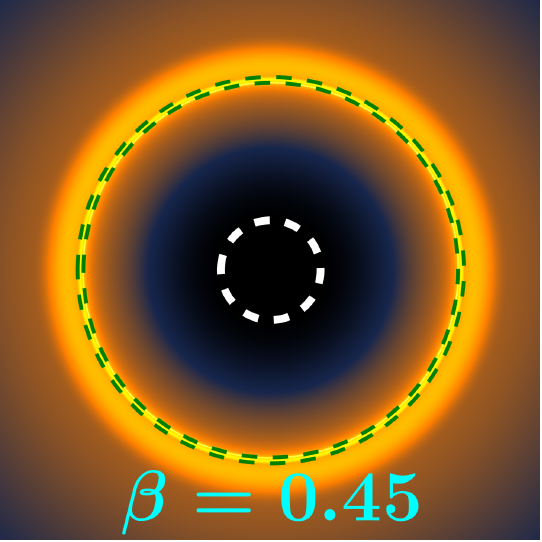}
\includegraphics[width=0.15\textwidth]{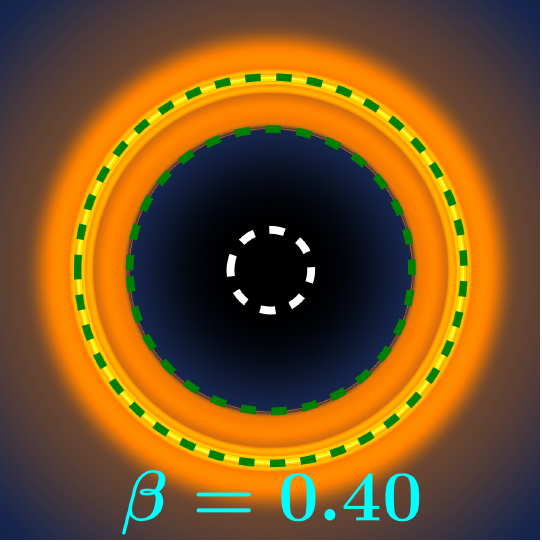}
\includegraphics[width=0.15\textwidth]{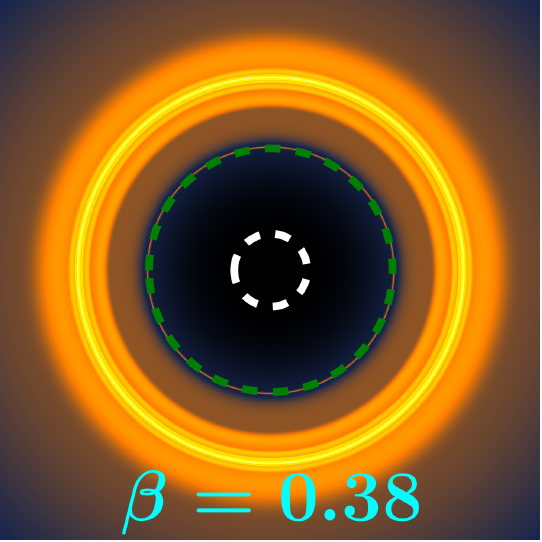}\\  
\caption{The photon ring (bright orange annulus) and shadow (central dark region) vary with the Kasner parameter $\beta$ for $Z=e^{0.9\psi^2}$. Green dashed circles mark unstable photon spheres, and white dashed circles denote inner shadow boundaries. Top panels show behavior near $q_c$, bottom panels far from $q_c$. Two photon spheres are observed for $\beta=0.45$ (bottom left) and $\beta=0.40$ (bottom middle), whereas only one photon sphere appears in the remaining cases.}\label{FigBetaDisk}
\end{figure}

The scaling law~\eqref{BetaToQ} provides a direct link between parameters characterizing the interior singularity and exterior observables such as the black hole charge, which motivates the search for external probes of black hole interiors. Direct imaging of black holes has become a powerful tool for probing their observable properties~\cite{EventHorizonTelescope:2019ths,EventHorizonTelescope:2019kwo}.
However, our ray-tracing simulations of black hole imaging suggest that such connections are not straightforwardly reflected in photon ring observations (see Fig.~\ref{FigBetaDisk} and SM). While the interior geometry undergoes drastic changes near $q_c$, the exterior shadow and photon ring structure remain nearly identical to those of the RN solution. Conversely, far from $q_c$, the exterior features exhibit rich variations even as the interior geometry stabilizes. This clear decoupling underscores the challenge of inferring black hole interior dynamics from astronomical observations. 
Nevertheless, we observe a critical scaling for the photon‑sphere radius $r_{\text{ph}}(q)$ near $q_c$: $r_{\text{ph}}(q)-r_{\text{ph}}(q_c)\sim \left(q-q_c\right)$, which follows the same scaling as the Kasner exponent $p_s$ in~\eqref{exponent}. Furthermore, for scalarized‑connected‑type models, the horizon value of the scalar field scales as $\psi(z_H)\sim (q-q_c)^{0.5}$, matching the scaling behavior of the Kasner exponent $p_\psi$ in~\eqref{exponent}.

\textbf{Conclusion.}--Our findings reveal fundamental aspects of black hole
interiors in EMS theories and offer novel perspectives on critical phenomena in gravitational systems. We establish a no-inner-horizon theorem, demonstrating that the spacetime ends in a spacelike Kasner singularity. This result strongly reinforces the strong cosmic censorship and ensures the internal predictability of charged black holes. Remarkably, we have disclosed a novel critical phenomenon for the emergence of Kasner singularity near the bifurcation point $q_c$ inside the black hole (see Fig.~\ref{FigCriticalBetaQ}). The observed scaling appears to be universal and not related to a specific choice of coupling function. It demonstrates that nonlinear regime of general relativity can produce unexpected, universal behavior and serves as a theoretical laboratory for understanding the dynamics of strong-field gravity. Our critical behavior emerges in the space of static solutions in the vicinity of a bifurcation point, with its imprint on the interior Kasner geometry. This stands in contrast to dynamical critical phenomena found in contexts like gravitational collapse (\emph{e.g.} type-I/II Choptuik-like scaling) or dynamical scalarization.
Particularly, it is distinct from the critical phenomena at the tip of the causal diamond within EMS black holes reported~\cite{Shao:2025fki}, which arises from gravitational collapse tuned by the initial profile of the scalar field. Thus, our study is { complementary} to exterior/dynamical studies of scalarization.

We interpret the Maxwell field $A_\mu$ of~\eqref{Action} broadly. It can be the usual electromagnetic field—astrophysical black holes may carry a small net { electric} charge (\emph{e.g.} Sgr A* $\le 3\times 10^8 C$~\cite{Zajacek:2018ycb}). The charge-to-mass ratio decreases monotonically as the coupling strength $\alpha^2$ increases and the hairy black holes satisfy observational bounds for sufficiently large $\alpha^2$~\cite{Herdeiro:2018wub}. {  Furthermore, electromagnetic duality maps the electrically charged black hole~\eqref{Ansatz} to a magnetically charged one, a description that is more astrophysically natural in magnetically dominated environments.} Alternatively, it can be a dark photon with dark charge, immune to plasma neutralization. This links black holes~\eqref{Ansatz} to dark matter via spontaneous scalarization, offering observable imprints at the intersection of modified gravity, dark matter, and multimessenger astronomy. Ultimately, any such scenario must be consistent with the full suite of observational constraints from gravitational waves, cosmology, black hole, and particle physics, see \emph{e.g.}~\cite{East:2022ppo,An:2024wmc,Caputo:2020bdy,Pierce:2018xmy}.  

Charged EMS models offer a useful setting for studying tachyonic instabilities and critical phenomena in scalarization, where the charge serves as a control parameter analogous to spin or curvature couplings. Notably, the universal interior scaling law may extend to more realistic scenarios, such as rotation-driven scalarization black holes. Therefore, several promising directions warrant further investigation. First, the universal critical behavior~\eqref{exponent} calls for an analytic derivation and deeper understanding. Second, it is compelling to investigate whether similar critical phenomena occur inside non-spherically-symmetric black holes and in other gravitational frameworks. Finally, although no direct connection has yet been established between imaging observables and interior dynamics in scalarized black holes, other external signatures may still serve as effective probes into their internal structure.

\section*{Acknowledgement}
We would like to thank Qian Chen, Yiqian Chen, and Zhuan Ning for helpful discussions. This work is supported by the National Key Research and Development Program of China Grant No.\,2021YFC2203004, and by the National Natural Science Foundation of China Grants No.\,12525503, No.\,12588101 and No.\,12447101. We acknowledge the use of the High Performance Cluster at the Institute of Theoretical Physics, Chinese Academy of Sciences.

\section*{End Matter}

\begin{figure}[h!]
\includegraphics[width=0.45\textwidth]{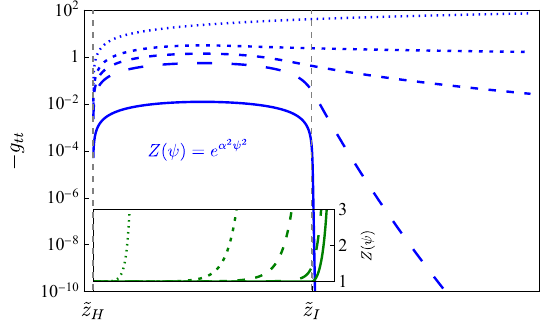}\\
\includegraphics[width=0.45\textwidth]{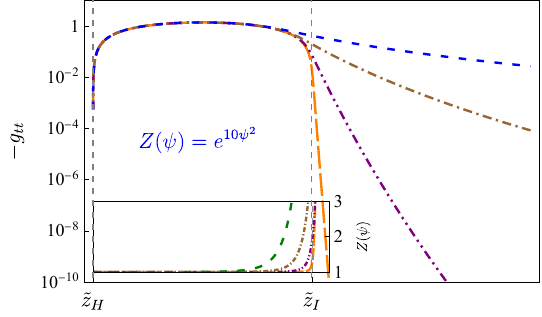}
\caption{The behavior of $g_{tt}$ near the would-be inner horizon (vertical dashed line) for $Z=e^{\alpha^2\psi^2}$. The horizontal axis is linearly scaled as $\tilde{z}=(z/z_H-1)/(z_I/z_H-1)$ so that the would-be inner horizon position $z_I$ corresponds to $\tilde{z}_I=1$, where $z_I$ is the inner horizon of the RN solution at $q=q_c$. \textbf{Top panel:} $q/q_c-1=10^{-6}$ fixed, with $\alpha^2=1,5,10,20,100$ (bottom to top).
\textbf{Bottom panel:} $\alpha^2=10$ fixed, with $q/q_c-1=10^{-6},10^{-7},10^{-8},10^{-9}$ (top to bottom). Insets show the coupling function $Z(\psi)$ as a function of $z$.}\label{FigCauchyHorizon}
\end{figure}

\textbf{Appendix A: Robustness of ER bridge collapse.}--Interestingly, we observe that tuning the coupling function $Z(\psi)$ leads to an apparent suppression of ER bridge collapse. This effect is demonstrated in the top panel of Fig.~\ref{FigCauchyHorizon} for the choice $Z=e^{\alpha^2\psi^2}$ with $\alpha^2$ a coupling constant. For our benchmark model at fixed $q/q_c$, ER bridge collapse is clearly present near the would-be Cauchy horizon when $\alpha^2$ is small. As the coupling parameter increases, however, the collapse weakens and eventually disappears for sufficiently large $\alpha^2$. Meanwhile, the inset shows that in this regime the nonlinearity of $Z$ becomes significant around the would-be Cauchy horizon—a feature also noted in a holographic superconductor model~\cite{Sword:2021pfm} (see also~\cite{Mirjalali:2022wrg}). Nevertheless, even for a sufficiently large $\alpha^2$, when $q$ is tuned close to $q_c$, the ER bridge collapse re-emerges distinctly, as illustrated in the bottom panel of Fig.~\ref{FigCauchyHorizon}. The inset further shows that the nonlinear structure of $Z$ near the would-be horizon becomes strongly suppressed as $q \to q_c$, for which one can follow the discussion~\cite{Hartnoll:2020rwq}. This confirms that ER bridge collapse is a robust manifestation of the inner horizon instability induced by the scalar degree of freedom (see also SM).

\textbf{Appendix B: Black hole spontaneous scalarization as a phase transition.}--Spontaneous scalarization can be understood as a continuous phase transition, well described by the phenomenological Landau model. In particular, the authors of~\cite{Muniz:2025egq} analyzed in detail a representative case involving neutron stars modeled as a perfect fluid within scalar-tensor theory, where scalarization is accurately captured as a continuous phase transition. They found that the critical exponents align with the standard predictions of Landau theory, and that a Landau-type ansatz provides a good fit for a suitably defined free energy near the critical point. We now demonstrate that a similar picture applies to the black hole case.

To draw a parallel with the case of scalarized neutron stars~\cite{Muniz:2025egq}, we now introduce a source for the scalar field at the black hole boundary by fixing $\psi_0 = \psi(z=0)$. Near the boundary, the scalar field behaves as
\begin{equation}
\psi(z) = \psi_0 + Q_\psi z + \cdots,
\end{equation}
where $Q_\psi$ denotes the scalar charge. The remaining boundary conditions does not change. With this setup, the hairy black holes now form a two-parameter family of inequivalent solutions, characterized by the charge-mass ratio $q$ and the scalar source $\psi_0$.

To frame spontaneous scalarization as a phase transition, we consider $q$ as determining the control parameter $\tau$. Specifically, we define $\tau = (q_c - q)/q_c$, where $q_c$ denotes the critical value. Following the approach in~\cite{Muniz:2025egq}, we identify the scalar charge $Q_\psi$ as the order parameter and the source $\psi_0$ as the external field. Then, the phase transition mechanism of black hole spontaneous scalarization is as follows. In the absence of an external field ($\psi_0 = 0$) and for $\tau > 0$, only one equilibrium solution exists, corresponding to $Q_\psi = 0$. Beyond the critical point ($\tau < 0$), the $Q_\psi = 0$ solution becomes unstable under scalar perturbations, and two new stable solutions emerge when the coupling preserves a $Z_2$ symmetry~\cite{Herdeiro:2018wub,Fernandes:2019rez}. This scenario closely mirrors the framework developed for scalarized neutron stars~\cite{Muniz:2025egq}.
\begin{figure*}[htp]
\centering
\includegraphics[width=0.32\textwidth]{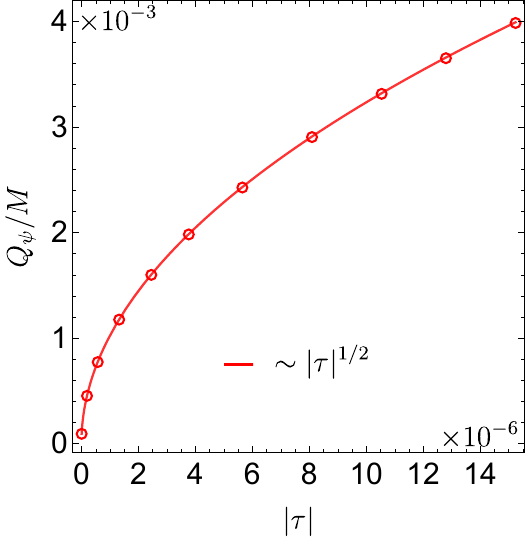}
\includegraphics[width=0.32\textwidth]{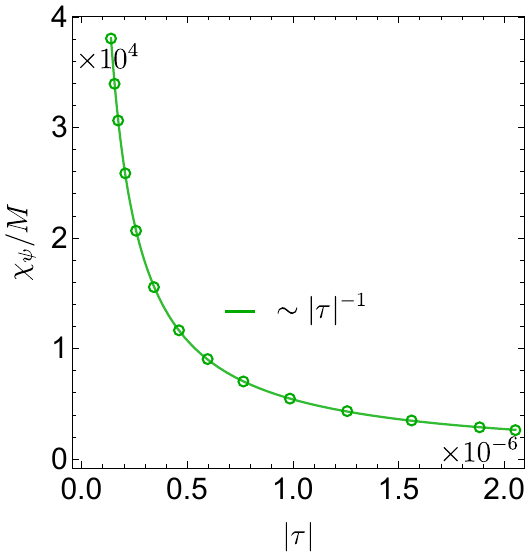}
\includegraphics[width=0.32\textwidth]{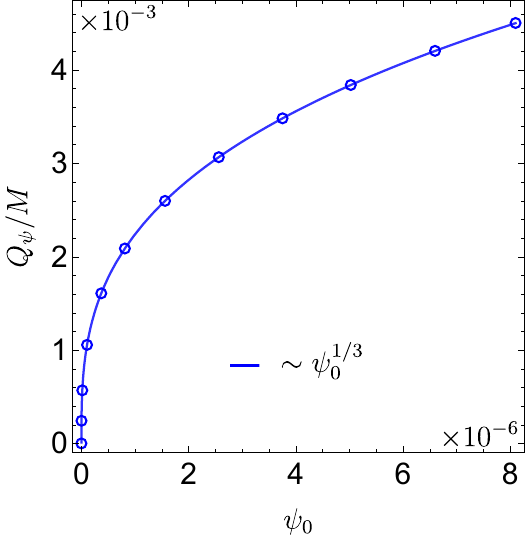}
\caption{Scaling behaviors for black hole spontaneous scalarization near the critical point $q_c$ with $Z(\psi)=1+\psi^2$.
\textbf{Left panel:} Scalar charge $Q_\psi$ versus the absolute value of the control parameter $\tau=(q_c-q)/{ q_c}$.
\textbf{Middle panel:} Scalar susceptibility $\chi_\psi$ as a function of $\tau$.
\textbf{Right panel:} Order parameter $Q_\psi$ versus external source $\psi_0$ at the critical point $q=q_c$.
Dots represent numerical data, and solid curves in all panels show best fits consistent with Landau theory predictions. We have worked in units of the black hole mass $M$.}\label{fig:S3}
\end{figure*}
\begin{figure}[htp]
\includegraphics[width=0.50\textwidth]{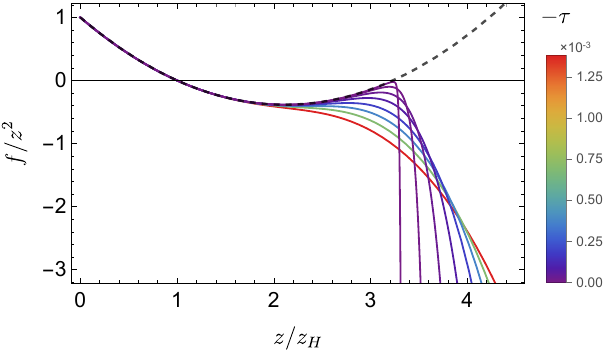}
\caption{The blackening factor { $f(z)/z^2$} both outside ($0<z<z_H$) and inside ($z>z_H$) the scalarized black hole for various values of control parameter $\tau=(q_c-q)/{ q_c}$. The black dashed line corresponds to the RN solution at $\tau=0$ ($q=q_c$), which vanishes at both the event and inner horizons. Near the critical point, the exterior geometry remains nearly indistinguishable from the RN solution, whereas the interior geometry undergoes pronounced deviations. To make the interior structure more visible, we take $Z = 1 + 4\psi^2$.}\label{Figfvsq}
\end{figure}

The critical exponents characterizing the phase transition can be computed explicitly within the Landau framework. To this end, we introduce the charge susceptibility, defined as
\begin{equation}
\chi_\psi = \left. \frac{\partial Q_\psi}{\partial \psi_0} \right|_{q},
\end{equation}
which measures the response of the order parameter to an external field. In the vicinity of the critical point $q_c$, the scalar field outside the horizon remains small, making it sufficient to approximate the coupling as $Z = 1 + \alpha^2 \psi^2$. As illustrated in Fig.~\ref{fig:S3}, the numerical solutions exhibit the following scaling behaviors near the critical point:
\begin{itemize}
    \item In the absence of an external field ($\psi_0 = 0$), the order parameter scales as $Q_\psi(q, \psi_0=0) \sim { |\tau|}^{1/2}$ as shown in the left panel of Fig.~\ref{fig:S3}.
    \item The zero-field susceptibility diverges as $\chi_\psi(q, \psi_0=0) \sim |\tau|^{-1}$, depicted in the middle panel of Fig.~\ref{fig:S3}.
    \item At the critical charge ratio ($q = q_c$), the order parameter responds nonlinearly to the external field, following $Q_\psi(q_c, \psi_0) \sim \psi_0^{1/3}$, as seen in the right panel of Fig.~\ref{fig:S3}.
\end{itemize}
These critical behaviors are fully consistent with the mean-field predictions of Landau's theory of continuous phase transitions, reinforcing the interpretation of spontaneous scalarization as an equilibrium critical phenomenon.

Near the critical point $q_c$, the exterior geometry remains nearly identical to the smooth RN solution. The interior geometry tells a strikingly different story: it undergoes drastic change near $q_c$, with the deformation becoming ever more severe as $q$ approaches the critical value. This is clearly visible in Fig.~\ref{Figfvsq} for the blackening factor { $f(z)/z^2$}. Our work thus uncovers a distinct critical scaling inside the horizon, revealing a novel interior manifestation of criticality absent from previous studies.

\section*{Supplementary material}

This supplementary material offers a detailed exposition of the analysis presented in the main text. We first derive the equations of motion and outline the numerical methodology for obtaining scalarized black hole solutions. We then examine the interior dynamics of these solutions, encompassing the collapse of the ER bridge and the subsequent emergence of Kasner epochs. Finally, we discuss the implications for black hole imaging.
\section{Equations of motion}\label{SectionSetup}
With varying the action in Eq.~(1) of the main text, the equations of motion for $g_{\mu\nu}$, $\psi$ and $A_{\mu}$ can be obtained as
\begin{equation}\label{EoM1}
\begin{aligned}
{\mathcal R}_{\mu \nu}-\frac{1}{2} {\mathcal R} g_{\mu \nu}=2\partial_\mu \psi \partial_\nu \psi -g_{\mu \nu}(\partial \psi)^2\\
+Z(\psi)\left(2F_{\mu \rho} {F_\nu}^\rho-\frac{1}{2} g_{\mu \nu} F^2\right)\,,\\
\nabla_{\mu} \nabla^{\mu} \psi=\frac{1}{4} \dot{Z}(\psi)F_{\mu \nu} F^{\mu \nu}\,,\\
\nabla_{\mu}\left(Z(\psi) F^{\mu \nu}\right)=0\,,
\end{aligned}
\end{equation}
with the dot denoting the derivative with respect to $\psi$. Substituting the ansatz  
\begin{equation}
\begin{aligned}
ds^2 = \frac{1}{z^2} \left[- f(z) \mathrm{e}^{-2\chi(z)} dt^2 + \frac{dz^2}{f(z)} +d\theta^2+\sin^2\theta d\varphi^2 \right],\\
\psi=\psi(z),\quad A_{\mu}{dx^\mu}=A_t(z)dt\,,
\end{aligned}
\end{equation}
into~\eqref{EoM1} yields the equations of motion
\begin{equation}\label{EoM2}
\begin{aligned}
\left(z^{-2}\mathrm{e}^{-\chi}f\psi'\right)'=-\frac{1}{2}\mathrm{e}^{\chi}\dot{Z}A_t'^2\,,\\
\left(\mathrm{e}^{\chi}ZA_t'\right)'=0\,, \quad \chi'=z\psi'^2\,,\\
\left(z^{-3}\mathrm{e}^{-\chi}f\right)'=-\frac{\mathrm{e}^{-\chi}}{z^2}+\mathrm{e}^{\chi}ZA_t'^2\,,\\
\end{aligned}
\end{equation}
where the prime denotes the derivative with respect to the radial coordinate $z$. For later convenience, we introduce a new function $h=z^{-3}\mathrm{e}^{-\chi}f$. Then, integrating the equation of motion about $A_t$ in~\eqref{EoM2}, we rewrite the above equations of motion as follows:
\begin{equation}\label{EoMQ}
\begin{aligned}
\psi''=-&\left(\frac{1}{z}+\frac{h'}{h}\right)\psi'+\frac{\mathrm{e}^{-\chi}Q^2}{2z h}\frac{\mathrm{d}}{\mathrm{d}\psi}\left(\frac{1}{Z}\right)\,,\\
\mathrm{e}^{\chi}&ZA_t'=-Q\,,\quad \chi'=z\psi'^2\,,\\
h'&=\mathrm{e}^{-\chi}\left(-\frac{1}{z^2}+\frac{Q^2}{Z}\right)\,,
\end{aligned}
\end{equation}
where $Q$ is the electric charge of the black hole. 

Given the nonlinear equations of motion~\eqref{EoMQ}, we have to solve them numerically, which requires appropriate boundary conditions both at the event horizon $z=z_H$ and the boundary $z=0$. Without loss of generality, we work in a gauge where $A_t$ vanishes on the event horizon. The smoothness of the geometry at the horizon admits the following horizon field expansions
\begin{equation}\label{zHBoundaryConditions}
\begin{aligned}
    h(z)& = h_1(z - z_H) + \cdots, \\
    \chi(z)&= \chi_H + \chi_1(z - z_H) + \cdots,\\
    \psi(z)&= \psi_H + \psi_1 (z - z_H) +\cdots,\\
    A_{t}(z)& = A_{t1}(z - z_H) +\cdots,
\end{aligned}
\end{equation}
with $(\chi_H, \psi_H, h_1, \chi_1, \psi_1, A_{t1})$ constants. Inserting these into~\eqref{EoMQ}, we find that~\eqref{zHBoundaryConditions} are fully determined by three parameters: $\psi_H$, $\chi_H$ and $A_{t1}$. As for the spatial infinity ($z=0$), asymptotic flatness imposes the boundary conditions $\chi(0) = \psi(0) = 0$. Moreover, the expansion of the equations at infinity reads
\begin{equation}\label{z0BoundaryConditions}
\begin{aligned}
z h(z) &=1-2Mz+\cdots,\\
A_t(z)&=\mu - Q z + \cdots,
\end{aligned}
\end{equation}
where $M$ and $\mu$ denote the ADM mass and electrostatic potential, respectively.

Noting that the equations of motion~\eqref{EoMQ} are invariant under the scaling
\begin{equation}\label{Scaling}
\begin{aligned}
(Q, M, h)\to \lambda (Q, M, h),\quad\ z \to \lambda^{-1} z\,,\\
(\psi, A_t,\chi)\to (\psi, A_t,\chi)\,,
\end{aligned}
\end{equation}
with $\lambda$ a constant. This is a scaling symmetry that relates different solutions. Therefore, once the coupling function $Z(\psi)$ is given, there is a one-parameter family of inequivalent hairy solutions labeled by the charge-to-mass ratio $q=Q/M$. Extending the numerical solutions into the interior of the black hole is straightforward. 

We highlight that in order to trigger a spontaneous scalarization with a purely electric field, the coupling should satisfy the condition that is
\begin{equation}\label{NoHorizonCond}
\psi \dot Z(\psi)>0, 
\end{equation}
for some range of $z$ outside the black hole by utilizing the properties of asymptotic flatness and the horizon~\cite{Astefanesei:2019pfq}. In practice, the condition~\eqref{NoHorizonCond} applies to a wide class of coupling functions commonly used in EMS models, such as
\begin{itemize}
    \item $Z=1+\alpha^2\psi^{2n}$,
    \item $Z=e^{\alpha^2\psi^{2n}}$ and $Z=\cosh(\sqrt{2}\alpha\psi^{n})$,
    \item $Z=1+\frac{\alpha^2\psi^{2n}}{1+\psi^{2n}}$,
\end{itemize}
where $n$ is a positive integer and $\alpha$ a constant. In the present study, we consider the models that satisfy~\eqref{NoHorizonCond}, thus the black hole interior generically ends at a space-like singularity as $z\rightarrow\infty$, in accordance with the no-inner-horizon result of the main text.

\begin{figure}[htp]
\includegraphics[width=0.45\textwidth]{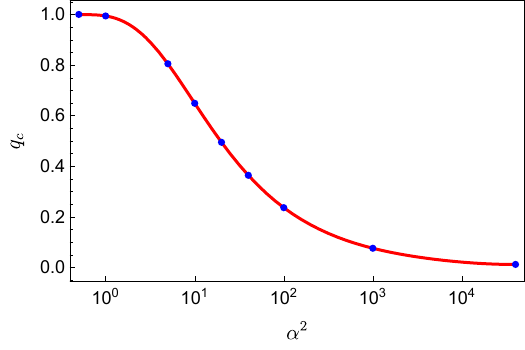}
\caption{Dependence of the critical charge-to-mass ratio $q_c$ on the coupling constant $\alpha^2$ for scalarized black hole formation. The value of $q_c$ decreases monotonically as $\alpha^2$ increases.}\label{Figqcalpha} 
\end{figure}

For scalarized black holes that bifurcate from the RN solution and reduce to it in the limit $\psi = 0$, the coupling function $Z(\psi)$ admits the small-$\psi$ expansion
\begin{equation}\label{eqZ}
Z(\psi) = 1 + \alpha^2 \psi^{2} +\cdots\,,
\end{equation}
where $\alpha^2$ denotes the coupling constant characterizing the strength of the scalar-Maxwell interaction. This quadratic form is sufficient to capture the tachyonic instability that triggers scalarization, provided the coupling strength exceeds the threshold value $\alpha^2 > 1/4$~\cite{Herdeiro:2018wub}. For a fixed $\alpha^2$ satisfying this condition, scalarized solutions exist beyond a critical charge-to-mass ratio $q_c$. As shown in Fig.~\ref{Figqcalpha}, the critical value $q_c$ decreases monotonically with increasing $\alpha^2$. Consequently, for sufficiently large coupling, scalarized black holes can be supported even for very small charge-to-mass ratios.

\section{Black hole interior dynamics}\label{SecInterior}
We have established that for a broad class of coupling functions satisfying condition~\eqref{NoHorizonCond}, scalarized black holes cannot possess smooth inner horizons, resulting in spacetime terminating at a spacelike singularity as $z\rightarrow\infty$. Here, we demonstrate that the scalar field triggers a rapid collapse of the ER bridge at the would-be inner Cauchy horizon, followed by the emergence of a Kasner geometry during the late-time interior evolution.

\subsection{ER bridge collapse}\label{ER}
The no-inner-horizon theorem reveals the instability of the inner horizon triggered by the scalar field. In the vicinity of the would-be Cauchy horizon, one anticipates the collapse of the ER bridge, for which, as the metric component $g_{tt}$ approaches its would-be zero value at the Cauchy horizon, it suddenly suffers a very rapid collapse and becomes exponentially small. The key fact is that for a very small scalar field, the instability is so fast that one can keep the $z$ coordinate essentially fixed and drop some terms from the equations of motion.

However, it has been argued that this collapse could be fully suppressed by strengthening the EMS coupling~\cite{Sword:2021pfm} where the key observation was that near the critical point, $g_{tt}$ does not significantly decrease near the would-be Cauchy horizon for sufficiently strong coupling. We find similar feature by tuning the coupling $Z(\psi)$, see the top panel of Fig.~2 in the main text and Fig.~\ref{FigCauchyHorizon2} below. 
For our benchmark models at a fixed $q/q_c$, we observe the collapse of the ER bridge around the would-be Cauchy horizon when the value of $\alpha^2$ is small. Nevertheless, by increasing the coupling parameter, the collapse appears to become less severe and does not occur for sufficiently large $\alpha^2$. As shown in the insert, the nonlinear nature of $Z$ is no longer negligible around the would-be Cauchy horizon. Nevertheless, if one adjusts $q$ so that it approaches the critical value closely, the behavior of ER bridge collapse can be clearly observed, see the bottom panels of Fig.~2 in the main text and Fig.~\ref{FigCauchyHorizon2}. Therefore, the ER bridge collapse should be a robust manifestation of the inner horizon instability induced by the scalar degree of freedom. By analogy, we expect that in the holographic superconductor model of~\cite{Sword:2021pfm}, adjusting the temperature toward its critical value should likewise restore a clear collapse of the ER bridge.

\begin{figure}[h!]
\includegraphics[width=0.45\textwidth]{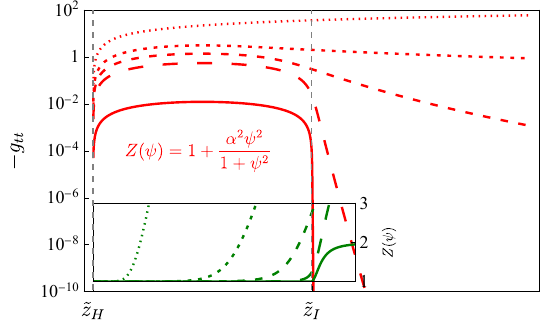}\\
\includegraphics[width=0.46\textwidth]{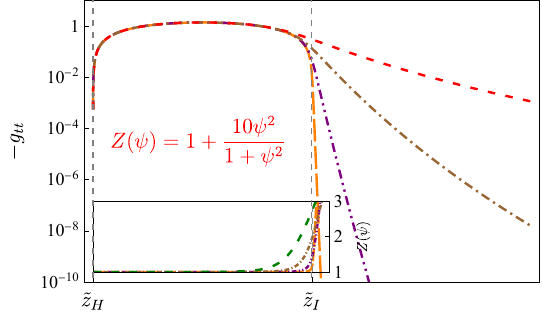}
\caption{The behaviors of $g_{tt}$ near the would-be Cauchy horizon (vertical dashed line) for $Z=1+\alpha^2\psi^2/(1+\psi^2)$. The horizontal axis has been linearly scaled as $\tilde{z}=(z/z_H-1)/(z_I/z_H-1)$, so that the position of the would-be Cauchy horizon $z_I$ always corresponds to $\tilde{z}_I=1$. Here $z_I$ is determined from the inner horizon of the RN solution at $q=q_c$. \textbf{Top panel:} Results for $q/q_c-1=10^{-6}$ with varying $\alpha^2$. Curves correspond (from bottom to top) to $\alpha^2=1, 5, 10, 20, 100$. \textbf{Bottom panel:} Results for fixed $\alpha^2=10$ with varying $q$. Curves correspond (from top to bottom) to $q/q_c-1=\{10^{-6},10^{-7},10^{-8},10^{-9}\}$. The inset in each panel displays the coupling function $Z(\psi)$ as a function of $z$. }\label{FigCauchyHorizon2}
\end{figure}

\subsection{Kasner singularity}\label{SubKasner}
Due to the highly nonlinear nature of the equations, analytic solutions are unavailable. Instead, following previous work~\cite{Cai:2023igv}, we seek self-consistent asymptotic forms near the singularity,
which will be further established by checking the full numerical solutions. 

%We are not able to solve the system analytically due to the strong nonlinear nature of the equations of motion. Following previous work~\cite{Cai:2023igv}, our strategy is to obtain self-consistent asymptotic solutions, which will be further established by checking the full numerical solutions. 

We begin with the assumption that the terms associated with the coupling $Z$ are all negligible in far interiors. Then, the approximate equations from~\eqref{EoMQ} are 
\begin{equation}\label{ApproximateEoM}
\begin{aligned}
\psi''=-\frac{1}{z}\psi'\,,\quad \chi'=z\psi'^2\,,\quad h'=-\frac{\mathrm{e}^{-\chi}}{z^2}\,.
\end{aligned}
\end{equation}
Since $\chi'$ is positive, $\chi$ is a monotonically increasing function. Therefore, we can conclude that $h'$ is integrable at this time because of
\begin{equation}
    \mathcal{O}(h')\leq\mathcal{O}\left(\frac{1}{z^2}\right).
\end{equation}
Hence, the solution~\eqref{ApproximateEoM} is given by
\begin{equation}\label{ApproximateSolution2}
\psi=\beta \ln z+C_{\psi}\,,\ \chi =\beta^2\ln z +C_{\chi}\,,\ h=C_h\,, 
\end{equation}
with $\beta$, $C_{\psi}$, $C_{\chi}$ and $C_{h}$ constants. With the coordinates transformation $\tau\sim z^{-\frac{3+\beta^2}{2}}$ from $z$ to proper time $\tau$, we obtain the Kasner geometry, the Eq.~(8) in the main text. More generally, a Kasner singularity in the presence of a scalar field $\psi$ is a spatially homogeneous but anisotropic cosmological singularity of the form $d s^2 =-d \tau^2 + \sum_{i=1}^d\tau^{2p_i} d x_i^2, \psi={-\frac{p_\psi}{\sqrt{2}}}\ln\tau$, with exponents satisfying $\sum_{i=1}^d p_i =p_\psi^2 +\sum_{i=1}^d p_i^2 = 1$~\cite{Belinskii:1973sud,Kasner:1921}.

Next, let's consider the contribution from the coupling function. In particular, we need to check if the terms we dropped are small in a given Kasner epoch.
We have the following properties for the coupling $Z$ as a function of the real scalar $\psi$:
\begin{equation}\label{ZConditions}
Z(\psi)\in\{Z(0)=1;\dot{Z}(0)=0;\psi\dot{Z}(\psi)>0\}\,.
\end{equation}
From the above condition, with the integral mean value theorem, one can easily find that
\begin{equation}
\begin{aligned}
Z(\psi)&=\int_0^\psi\dot{Z}(s)\mathrm{d}s+Z(0)\,,\\
&=\frac{\psi}{\psi_m} \psi_m\dot{Z}(\psi_m)+Z(0)\,,\\
&>\psi_m\dot{Z}(\psi_m)+1>1\,,
\end{aligned}
\end{equation}
where $\psi_{m}$ lies between 0 and $\psi$. Therefore, the coupling function $Z$ is positive and lower bounded with $Z(0)=1$ and $1/Z$ is also a bounded function, \emph{i.e.} $0<1/Z\leq1$. For sufficiently large $|\psi|$, together with~\eqref{NoHorizonCond}, it must be that
\begin{equation}
    \mathcal{O}\left(\frac{\mathrm{d}}{\mathrm{d}\psi}\left(\frac{1}{Z}\right)\right)<\mathcal{O}\left(\frac{1}{\psi}\right),
\end{equation}
in which we assume that the order of the {$1/Z$} can have a uniform upper bound as $\psi\to\infty$. In the equations of motion~\eqref{EoMQ}, with approximate solution~\eqref{ApproximateSolution2}, one can estimate that
\begin{equation}
\begin{aligned}
 \mathcal{O}\left(\frac{e^{-\chi}}{Z}\right) \leq\mathcal{O}\left(\frac{1}{z^{\beta^2}}\right),\\  \mathcal{O}\left(\frac{\mathrm{e}^{-\chi}}{z h}\frac{\mathrm{d}}{\mathrm{d}\psi}\left(\frac{1}{Z}\right)\right)<\mathcal{O}\left(\frac{1}{z^{\beta^2+1}\ln z}\right)\,.
\end{aligned}
\end{equation}
When $|\beta|>1$, the neglected terms for $Z$ will not change the Kasner dynamics, and thus the Kasner solution~\eqref{ApproximateSolution2} is stable.

The other case with $|\beta|<1$ is more complicated, for which the coupling function $Z$ could play an important role and a new Kasner solution might develop. 
The dynamics at this point are highly sensitive to the choice of the specific coupling function $Z(\psi)$. Providing a completely systematic understanding is beyond the scope of this work. However, we attempt to outline some general properties. We begin with introducing the following change in variables
\begin{equation}\label{AlphaPsi}
  \psi=\int^z\frac{\beta(s)}{s} ds\,.
\end{equation}
Then, the approximate equations of motion now become
\begin{equation}\label{EoMZBeta}
\begin{aligned}
\frac{\beta'}{\beta}&=-\frac{Q^2\mathrm{e}^{-\chi}}{h}\left(\frac{1}{Z}+\frac{1}{2 \beta }\frac{\mathrm{d}}{\mathrm{d}\psi}\left(-\frac{1}{Z}\right)\right)\,,\\ 
h'&=\frac{Q^2\mathrm{e}^{-\chi}}{Z},\quad \chi'=z\psi'^2\,.
\end{aligned}
\end{equation}
Here we have dropped the first term in the brackets of the last equation of~\eqref{EoMQ}; otherwise, the coupling function $Z$ would not play a role in that expression. It means that we have $\mathcal{O}(Z)<\mathcal{O}{(z^2)}$.

Given that $\psi$ is unbounded when $z\gg z_H$. One then finds that the sign of $\psi$ will eventually be the same as $\beta(z)$ in~\eqref{AlphaPsi}, from which one has 
\begin{equation}
\begin{aligned}
\frac{1}{2 \beta }\frac{\mathrm{d}}{\mathrm{d}\psi}\left(-\frac{1}{Z}\right)=\frac{1}{2 Z^2\beta^2 }\beta \dot{Z}>0\,.
\end{aligned}
\end{equation}
Note also that $h<0$ inside the hairy black hole, thanks to the no-inner-horizon theorem in the main text. 
Together with the first equation of~\eqref{EoMZBeta}, one finds that 
\begin{equation}\label{BetaConstraint}
    \beta'\beta>0\,.
\end{equation}
Thus, as approaching the singularity, the function $|\beta(z)|$ will increase monotonically. It explains the increase of $z\psi'(z)$ shown in Figure 4 of~\cite{Dias:2021afz}. Nevertheless, whether there will develop a second Kasner epoch depends on the details of $Z(\psi)$. For the Kasner epoch with an exponential-like coupling, \emph{e.g.} $Z\sim e^{\psi^{2}}$, even when its Kasner parameter $|\beta|<1$, both terms from $Z$ in~\eqref{EoMQ} are suppressed significantly, for which one still has a stable Kasner epoch. For the coupling with a power-law form or it has an upper bound, $h'$ from~\eqref{EoMZBeta} is not integrable when the exponent $|\beta|<1$. Therefore, the integral of $h'(z)$ yields $h(z)\rightarrow \infty$ as approaching the singularity, which is not possible as the no-inner-horizon theorem requires $h<0$ inside the event horizon. As a consequence, new dynamics will come into play, triggering the transformation to another epoch. The transition from a Kasner epoch with $|\beta|<1$ to a stable Kasner epoch with $|{\beta}|>1$ was observed inside an asymptotically flat black hole with charged scalar hair~\cite{Dias:2021afz}, which corresponds to our case with $Z=1+\alpha^2\psi^2$. Note that the exponent defined in~\cite{Dias:2021afz} is related to our present work by $\beta_{\text{there}}=\beta/\sqrt{2}$. Importantly, this additional Kasner alternation does not alter the scaling behavior reported in the main text in the vicinity of the critical point, where $|\beta| \gg 1$ holds.

\section{Black hole imaging with Kasner interiors}\label{SecBlackHoleImaging}
Based on the critical case where a relationship between the interior parameter $\beta$ and the exterior observable $q$ was established, this section explores whether such a connection can be generalized. We aim to determine if an external observable linked to the black hole’s interior structure can be identified in broader scenarios, thereby providing a potential probe for examining black hole interiors.

Thanks to the remarkable progress in black hole imaging, most notably the EHT observations of supermassive black holes~\cite{EventHorizonTelescope:2019ths,EventHorizonTelescope:2019kwo}, direct black hole images have become a powerful tool to probe their observable properties. Such images typically feature two main components: the shadow and the photon ring. %(bright ring corresponding to the unstable photon sphere). (central dark region) (bright ring formed by highly lensed photons)
The presence of the scalar field can, in certain regimes, significantly alter the photon sphere radius and the observational signature of the image~\cite{Gan:2021xdl,Al-Badawi:2024dzc,Wu:2025hcu}, opening a new window to test general relativity in the strong field regime. For instance, for the EMS model with exponential coupling function studied in~\cite{Gan:2021xdl}, double unstable photon spheres and broad photon ring band can appear in specific parameter ranges.

The previous sections have examined the interior structure of EMS scalarized black holes, characterized by the absence of inner horizons and the emergence of a Kasner spacetime controlled by the parameter $\beta$. We now study whether such interior structures have any manifestation in observable features with black hole imaging. In particular, we consider unstable photon spheres which play an important role in determining the accretion disk image seen by a distant observer. 

%optically and geometrically thin 
We consider the observational appearance of an accretion disk around a hairy black hole. It is convenient to work with the new coordinate $r=1/z$ for which the hairy black hole, the Eq.~(2) of the main text, becomes
\begin{equation}
ds^2=-N(r)e^{-2\chi(r)} dt^2+\frac{dr^2}{N(r)}+r^2d\Omega^2_2\,,
\end{equation}
where $N(r)=f(z)/z^2$. The unstable photon sphere is determined by the effective potential
\begin{equation}
    V_{\text{eff}}(r) = \frac{N(r) e^{-2{\chi(r)}}}{r^2}\,,
\end{equation}
with the following conditions~\cite{Gan:2021xdl}:
\begin{equation}
 V_{\text{eff}}(r_{\text{ph}})=\frac{1}{b_{\text{ph}}^2}, \;  V_{\text{eff}}'(r_{\text{ph}})=0,\;  V_{\text{eff}}''(r_{\text{ph}})<0\,.    
\end{equation}
Here $r_{\text{ph}}$ and $b_{\text{ph}}$ are the radius of the photon sphere and the corresponding impact parameter, respectively.

Assuming an optically and geometrically thin accretion disk in the equatorial plane, we employ backward ray tracing to obtain the corresponding black hole images. 
These images are characterized by several distinct features~\cite{Gralla:2019a,Chael:2021inner}: the standard shadow, the dark region enclosed by the (smaller) photon sphere; the inner shadow, corresponding to rays that fall into the event horizon before crossing the plane of the disk; and the photon ring, a bright annulus formed by light rays that intersect the plane of the disk at least three times.
Fig.~4 in the main text shows the results for the model with $Z(\psi)=e^{0.9 \psi^2}$. The top row of Fig.~4 illustrates behavior near the critical point $q_c$, where both the shadow and photon ring of the scalarized black hole remain nearly identical to those of the RN solution. In contrast, within our model, a slight variation of $q$ near $q_c$ triggers a pronounced collapse of the ER bridge in the interior, followed by a transition to a Kasner geometry with a significantly large $\beta$ that also changes drastically
(see the green dotted curve in the top panel of Fig.~2).

\begin{figure}[h!]
\includegraphics[width=0.45\textwidth]{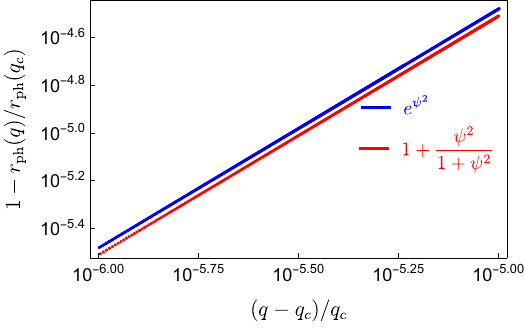}
\caption{The critical behavior between the radius of the photon sphere and the charge-to-mass ratio $q$ of {scalarized} black holes. Two examples are considered respectively.}\label{FigCriticalRph}
\end{figure}

The bottom row of Fig.~4 displays images of hairy black holes further away from $q_c$. Here, the interior geometry remains almost unchanged—the Kasner parameter $\beta$ varies only gradually—while the exterior structure shows considerable variation. The cases with 
$\beta=0.45$ (bottom left) and $\beta=0.40$ (bottom middle) both exhibit two unstable photon spheres, though in the former they lie very close to each other. The final case (bottom right), however, possesses only a single photon sphere. Our analysis indicates that the drastic near-threshold variation in the interior Kasner parameter $\beta$ does not produce correspondingly dramatic changes in the exterior photon sphere or shadow near $q_c$. Instead, the photon-sphere radius exhibits a simple linear scaling 
\begin{equation}
r_{\text{ph}}(q)-r_{\text{ph}}(q_c)\sim \left(q-q_c\right)\,,   
\end{equation}
analogous to but distinct from the interior $\beta$ scaling. Two examples are shown in Fig.~\ref{FigCriticalRph} respectively. Other scalarized-connected-type models have the same scaling behavior.

\bibliography{refs.bib}

\end{document}